\documentstyle[11pt,epsf]{article}
\newcommand{\al}{\alpha}

\newcommand{\bv}{\begin{verbatim}}

\newcommand{\ch}{\chi}

\newcommand{\ep}{\epsilon}
\newcommand{\et}{\eta}
\newcommand{\ev}{\end{verbatim}}
\newcommand{\fr}{\frac}
\newcommand{\Ga}{\Gamma}

\newcommand{\na}{\nabla}

\newcommand{\ph}{\phi}

\newcommand{\rh}{\rho}

\newcommand{\si}{\sigma}

\newcommand{\be}{\begin{equation}}
\newcommand{\ee}{\end{equation}} 
\newcommand{\eei}{\end{equation}\indent\indent}
\newcommand{\bc}{\begin{center}}
\newcommand{\ec}{\end{center}}
\newcommand{\ber}{\begin{eqnarray}}
\newcommand{\ear}{\end{eqnarray}}
\newcommand{\ba}{\begin{array}}
\newcommand{\ea}{\end{array}}

\newcommand{\p}{\partial}

\def\case#1/#2{\textstyle\frac{#1}{#2} }
\begin{document}
\title{The Physical Interpretation of the Lanczos Tensor.}
\author{Mark D. Roberts, \\\\
Department of Mathematics and Applied Mathematics, \\ 
University of Cape Town,\\
Rondbosch 7701,\\
South Africa\\\\
roberts@gmunu.mth.uct.ac.za} 
\date{\today}
\maketitle
\vspace{0.1truein}
\bc Published:  {\it Il Nuovo Cim.} {\bf B110}(1996)1165-1176.\ec
\bc Eprint: gr-qc/9904006\ec
\bc Comments:  15 pages, no diagrams,  no tables,  Latex2e.\ec
\bc 3 KEYWORDS:\ec
\bc Lanczos Tensor:~~  
    Aharonov-Bohm effect:~~  
    Quantum Potentials.\ec
\bc 1999 PACS Classification Scheme:\ec
\bc http://publish.aps.org/eprint/gateway/pacslist \ec
\bc 04.60+n,  03.65-w  .\ec
\bc 1991 Mathematics Subject Classification:\ec
\bc http://www.ams.org/msc \ec
\bc 83C45,  81S99,  81T13\ec
\newpage
\begin{abstract}
The field equations of general relativity can be written as first
order differential equations in the Weyl tensor,  the Weyl tensor in turn
can be written as a first order differential equation in a three index 
tensor called the Lanczos tensor.   Similarly in electro-magnetic theory 
Maxwell's equations can be written as first order differential equations in 
the field tensor $F_{ab}$ and this in can be written as a first order 
differential equation in the vector potential $A_a$;  
thus the Lanczos tensor plays a similar role in general relativity 
to that of the vector potential in electro-magnetic theory.   
The Aharonov-Bohm effect shows that when quantum mechanics is applied
to electro-magnetic theory the vector potential is dynamically significant,  
even when the electro-magnetic field tensor $F_{ab}$ vanishes.   
Here it is assumed that in the quantum realm the Lanczos tensor 
is dynamically significant,  and this leads to an attempt to quantize 
the gravitational field by pursuing the 
analogy between the vector field and the Lanczos tensor.
\end{abstract}
\section{Introduction}
\label{sec:intro}
The field equations of general relativity are usually written in terms of 
the Einstein tensor and the stress tensor;  however there is an alternative 
expression,  called Jordan's formulation of the field equations,  in which the 
field equations are expressed as first order equations in the Weyl tensor 
\cite{bi:HE}.
The Weyl tensor can be expressed,  using first order equations,  in terms of a
three index tensor 
\cite{bi:lanczos}
\cite{bi:mdr88}
\cite{bi:mdr89}
\cite{bi:mdr92} 
\cite{bi:mdr94} called the Lanczos tensor.
This tensor can be used to produce gravitational energy tensors of the
correct dimension \cite{bi:mdr88},  and these can be used to measure the speed
of gravitational waves \cite{bi:mdr94}.   
In Jordan's formulation the field equations are of a similar form 
to the Maxwell equations in terms of the electro-magnetic field tensor.   
The electro-magnetic field tensor can be expressed as a first order 
differential equation in the vector potential,  and thus the Lanczos 
tensor in analogous to the vector potential in electro-magnetic theory.   
The Lanczos tensor is not the only tensor that can be thought
of as being analogous to the vector potential;  because the field equations are
second order in the metric it is possible also to think of the metric (or the
difference between the metric and the Minkowski metric) as being analogous to 
the the vector potential.   There is also the Ashtekar potential in the theory 
of Ashtekar variables \cite{bi:ashtekar};  
this potential is not the same object as the Lanczos tensor,  
because the equation for the Weyl tensor involves cross terms 
in the Ashtekar potential,  unlike the Lanczos tensor in which 
the expression for the Weyl tensor is linear.   
The differential equations involving the Lanczos tensor which govern the 
dynamics of the field equations do not have a Lagrangian formulation,  thus 
traditional methods of quantization cannot be applied to the field equations 
in this form.

In electro-magnetic theory the vector potential was first introduced in 
order to express the equations of classical electrodynamics in simpler form.   
In classical physics the only physical effect of an electro-magnetic field 
on a charge is the Lorentz force,  and this only exists in regions where the 
electric or magnetic field in non-vanishing.   
The Aharonov-Bohm effect 
\cite{bi:ES}
\cite{bi:AB}
\cite{bi:BH}
demonstrates that this is not so in quantum mechanics;  physical effects occur 
in regions where the electric and magnetic fields both vanish,  but where the 
vector potential does not vanish.   It has been experimentally confirmed 
\cite{bi:chambers}.

In general relativity the existence of th Lanczos tensor might be just a 
technical curiosity,  or it might have fundamental significance in the way that
the vector potential does in electro-magnetic theory.   The object of the 
present paper is to determine a thought experiment,  similar to the Aharonov-
Bohm experiment,  which would in principle determine whether the Lanczos tensor
effects the dynamics and so is physically significant.   At the quantum level 
the vector potential enters the Schrodinger equation through the application 
of the electro-magnetic covariant derivative.   
The main problem in our approach 
is what should correspond to this covariant derivative;  after all there is 
already a covariant derivative in general relativity constructed from the 
Christoffel symbol.   Here it is postulated that in the quantum realm a 
covariant derivative involving the Lanczos tensor plays a role,  and that it is
the correct covariant derivative to apply to the Schrodinger equation in 
analogs of the Aharonov-Bohm effect.   
Now the main problem becomes how precisely 
should this covariant derivative be constructed;  
this cannot be know {\it a-priori} 
and so it is necessary to construct an example which will motivate a suitable 
definition of a new covariant derivative.  If such a covariant derivative could
be constructed it is possible to anticipate several difficulties.   
{\it Firstly} why
does the classical theory have no use for a Lanczos covariant derivative?   
{\it Secondly} in the quantum realm the Christoffel covariant derivative 
is still necessary to connect the Lanczos tensor to the Weyl tensor,  
why have two covariant derivatives?   
{\it Thirdly},  in the Aharonov-Bohm effect the electro-magnetic field 
is not quantized,  only the test particles are treated in a quantum 
mechanical manner;  in our case would only the test particles be treated
quantum mechanically,  or would the Lanczos tensor or the metric or both
be quantum fields?

The Aharonov-Bohm effect depends crucially on the existence of a choice of
the vector potential which is well-defined and continuous everywhere.   
For example,   as discussed in \cite{bi:BH},  
it is possible to choose a gauge in which 
the vector potential vanishes outside the solenoid and claim that there should
be no effect;  in fact,  as experiments \cite{bi:chambers} vindicate,  
a gauge should be chosen in which the vector potential is continuous 
everywhere.   It is assumed 
that a similar criteria on continuity exists for the present examples,  so that
it is possible to fix the gauge and then calculate the global effect of 
having a non-vanishing Lanczos tensor in the exterior region of the space-time.
In the examples presented here it is found that the Lanczos tensor is either 
continuous or not depending on whether the derivative of the metric is 
continuous or not,  irrespective of the choice of gauge.   
Thus there is no criteria to inform
us which is the correct gauge,  and hence the analogy cannot be carried through
completely.   This might be because the present examples are so geometrically 
simple.   The Aharonov-Bohm effect depends on a current in a solenoid and it is
not clear what is the general relativistic analog of a current in a solenoid.  
The simple cylindrical space-time used here might just be analogous to a line 
of charges,  for which the Aharnov-Bohm effect does not work.   Perhaps the 
correct analogy is a fluid in a pipe,  i.e. a cylindrical space-time with a 
perfect fluid moving along the axis;  
however such an exact solution is not known.

The example discussed here use a simple cylindrically symmetric space-time.
This space-time is general enough to include the simplest cosmic string 
\cite{bi:vilenkin} \cite{bi:gott}.
The approach used here relies on the Aharonov-Bohm effect being a quantum
mechanical effect,  and should not be confused with classical analogs 
of the Aharonov-Bohm effect which exist in cosmic string and some other 
space-times \cite{bi:bezerra}.

In section \ref{sec:II} the elementary properties of the 
Lanczos tensor are expounded.
In section \ref{sec:III} the Lanczos tensor is produced 
for a simple cylindrically symmetric space-time.   
In section \ref{sec:IV} the construction of covariant derivatives
involving the Lanczos tensor are discussed.
\section{The Lanczos Tensor.}
\label{sec:II}
The field equations of general relativity can be re-written in Jordan's 
form \cite{bi:HE}
\ber
&&C^{~~~d}_{abc.;d}=J_{abc},\nonumber\\
&&J_{abc}=R_{ca;b}-R_{cb;a}+\fr{1}{6}g_{cb}R_{;a}-\fr{1}{6}g_{ca}R_{;b},
\label{eq:1}
\ear
which is analogous to Maxwell's equations
\be
F^{~b}_{a.;b}=J_a.
\label{eq:2}
\ee
The Weyl tensor can be expressed in terms of the Lanczos tensor
\cite{bi:lanczos}
\cite{bi:mdr88}
\cite{bi:mdr89}
\cite{bi:mdr92}
\ber
C_{abcd}&=&H_{abc;d}-H_{abd;c}+H_{cda;b}-H_{cdb;a}\nonumber\\
&-&(g_{ac}(H_{bd}+H_{db})-g_{ad}(H_{bc}+H_{cb})+
    g_{bd}(H_{ac}+H_{ca})-g_{bc}(H_{ad}+H_{da}))/2\nonumber\\
&+&2H^{ef}_{..e;f}(g_{ac}g_{bd}-g_{ad}g_{bc})/3,
\label{eq:3}
\ear
where the Lanczos tensor has the symmetries
\be
H_{abc}+H_{bac}=0,\\
\label{eq:4}
\ee
\be
H_{abc}+H_{bca}+H_{cab}=0,  
\label{eq:5}
\ee
and where $H_{bd}$ is defined by
\be
H_{bd}\equiv H^{~e}_{b.d;e}-H^{~e}_{b.e;d}.
\label{eq:6}
\ee
Equation \ref{eq:3} is invariant under the algebraic gauge transformation
\be
H_{abc}\rightarrow H'_{abc}=H_{abc}+\ch_a g_{bc}-\ch_b g_{ac},
\label{eq:7}
\ee
where $\ch_a$ is an arbitrary four vector.

The Lanczos tensor with the above symmetries has 20 degrees of freedom,
but the Weyl tensor has 10.   
Lanczos \cite{bi:lanczos} reduced the degrees of freedom to 10 
by choosing the Lanczos algebraic gauge
\be
3\ch_a=H^{~b}_{a.b}=0,
\label{eq:8}
\ee
and the Lanczos differential gauge
\be
L_{ab}=H^{~~c}_{ab.;c}=0.
\label{eq:9}
\ee
These gauge choices are in some ways different than those in electro-magnetic
theory.   The algebraic gauge is different because it is algebraic and not 
differential in nature.   The differential gauge is different because a 
differential gauge transformation alters components which do not participate
in constructing the Weyl tensor;  in electro-magnetic theory a gauge 
transformation alters components in the vector potential all of which 
participate in constructing the electro-magnetic tensor.   
These difference are well illustrated by the example in the next section.

When the Lanczos tensor happens to be the gradient of an anti-symmetric
tensor of the second order
\be
H_{abc}=F_{ab;c},
\label{eq:10}
\ee
and if the Lanczos tensor is in the algebraic gauge \ref{eq:8}, \ref{eq:5}, 
and \ref{eq:10} imply that $F_{ab}$ obeys Maxwell's equations.   
It is not possible to introduce a source $J_a$ to \ref{eq:10} 
without it having an un-natural constraint by virtue of the
identity \ref{eq:3}.

In the case of weak gravity 
\be
g_{ab}=\et_{ab}+h_{ab},
\label{eq:11}
\ee
where $\et_{ab}$ is the Minkowski metric and $h_{ab}$ 
and its derivatives are small,  the Lanczos tensor can be written as
\be
4H_{abc}=\p_b h_{ac}-\p_a h_{bc}
         +\fr{1}{6}h_{,a}\et_{bc}-\fr{1}{6}h_{,b}\et_{ac},
\label{eq:12}
\ee
where $h=h^a_{.a}$.
\section{The Lanczos Tensor for a Simple Cylindrically Symmetric Space-time.}
\label{sec:III}
In this section we find the Lanczos tensor for a simple static 
cylindrically symmetric space-time with line element
\be
ds^2=-dt^2+dr^2+X~d\ph^2+dz^2.
\label{eq:13}
\ee
The non-vanishing Christoffel symbols are
\be
\Ga^r_{\ph\ph}=-\fr{1}{2}X_r,~~~
\Ga^\ph_{\ph r}=X_r /2X.
\label{eq:14}
\ee
The Riemann,  Ricci,  Einstein,  and Weyl tensors are conveniently 
expressed in terms of the Ricci scalar
\be
R=2R^\ph_{.\ph}=2R^r_{.r}
 =-2G^t_{.t}=-2G^z_{.z}
 =2R^\ph_{.r\ph r}=2R^r_{.\ph r\ph}/X,
\label{eq:15}
\ee
\be
R=-6C_{tztz}=6C_{\ph r\ph r}/X=12C_{t\ph t\ph}/X=-12C_{z\ph z\ph}/X 
 =12C_{t\ph t\ph}=-12C_{trtr},
\label{eq:16}
\ee
[note added 1999 which $C_{t \ph t\ph}$ is correct]
where
\be
R=-_{rr}X/X+X^2_r/2X=-X^{-\fr{1}{2}}(X'X^{-\fr{1}{2}})'.
\label{eq:17}
\ee

This space-time is general enough to include the simple cosmic string for 
which the metric is \cite{bi:vilenkin} \cite{bi:gott}
\be
ds^2=-dt^2+dr^2+\rh*sin^2(\rh/\rh^*)d\ph^2+dz^2,
\label{eq:18}
\ee
where $\rh^*=(8\pi\ep)^\fr{1}{2}$ and $\ep$ is the density of the string.   
The Ricci scalar is
\be
R=4\pi\ep.                                                            
\label{eq:19}
\ee
At the join $r=r_0$, $\rh=\rh_0$ the interior metric is attached 
to the exterior metric
\be
ds^2=-dt^2+dr^2+a^2r^2d\ph^2+dz^2,
\label{eq:20}
\ee
where $a$ is given by
\ber
&&a=1-4\mu,\nonumber\\
&&\mu=\int^{\rh_0}_0\int^{2\pi}_0\ep\rh^*sin(\rh/\rh^*)d\ph d\rh
     =2\pi\rh^{*2}(1-cos(\rh/\rh^*)),
\label{eq:21}
\ear
and $\mu$ is called the linear energy density.   The requirement that the 
metric is continuous at the join is
\be
ar_0=\rh^* sin(\rh/\rh^*).
\label{eq:22}
\ee
The derivative of the metric is continuous at the join,  as this is required 
for there to be no surface stress present;  this requirement gives a in 
\ref{eq:20},  otherwise a would be simply absorbed into the line element.

From \ref{eq:3} and \ref{eq:15} we have from the $C_{tztz}$ 
or $C_{\ph r\ph r}$ component
\be
R=2\p_rH_{trt}-2\p_rH_{\ph r\ph}+4\p_rH_{\ph r\ph}/X 
   +X_rH_{rtr}/X-X_rH_{zrz}/X-2X_rH_{\ph r\ph}/X^2,
\label{eq:23}
\ee
from the $C_{t\ph t\ph}$ or $C_{trtr}$ component
\be
R=-4\p_rH_{trt}-8\p_rH_{zrz}+4\p_rH_{\ph r\ph}/X +4X_rH_{trt}/X
  +2X_rH_{zrz}/X-2X_rH_{\ph r\ph}/X^2,
\label{eq:24}
\ee
from the $C_{\ph z\ph}$ or $C_{trtr}$ component
\be
R=8\p_rH_{trt}+4\p_rH_{zrz}+4\p_rH_{\ph r\ph}/X-2X_rH_{trt}/X
  -4X_rH_{zrz}/X-2X_rH_{\ph r\ph}/X^2,
\label{eq:25}
\ee
Subtracting \ref{eq:24} from \ref{eq:23} or \ref{eq:25} from \ref{eq:24} 
we have
\ber
0&=&2\p_rH_{trt}+2\p_rH_{zrz}-X_rH_{trt}/X-X_rH_{zrz}/X\nonumber\\
 &=&2\sqrt{X}\p_r(H_{trt}X^{-\fr{1}{2}})+2\sqrt{X}\p_r(H_{zrz}X^{-\fr{1}{2}}),
\label{eq:26}
\ear
which integrates to give
\be
H_{trt}=k\sqrt{X}-H_{zrz},
\label{eq:27}
\ee
where $k$ is a constant.   From \ref{eq:27} and \ref{eq:23} or \ref{eq:24} 
or \ref{eq:25} we have
\be
R=2kX_rX^{-\fr{1}{2}}-4\p_rH_{zrz}+4\p_rH_{\ph r\ph}/X -2X_rH_{zrz}/X
  -2X_rH_{\ph r\ph}/X^2.
\label{eq:28}
\ee
Here \ref{eq:8} the Lanczos algebraic condition is
$-H_{trt}+H_{zrz}+H_{\ph r\ph}/X=0$,  it gives
\be
H_{zrz}=k\sqrt{X}/2-H_{\ph r\ph}/2X.
\label{eq:29}
\ee
Equations \ref{eq:17},  \ref{eq:28},  and \ref{eq:29} give
\be
R=-3X_rH_{r\ph r}/X+6\p_rH_{\ph r\ph}/X=-X_{rr}/X+X_r^2/2X^2,
\label{eq:30}
\ee
integrating
\be
H_{\ph r\ph}=-X_r/6+l\sqrt{X}/6,
\label{eq:31}
\ee
where $l$ is a constant.   From ref{eq:27},  \ref{eq:28},  and \ref{eq:31} 
and inserting the gauge vector we have
\ber
&&H=\fr{k\sqrt{X}}{2}+\fr{l}{12\sqrt{X}}-\fr{X_r}{12X}+\ch_r,\nonumber\\
&&H=\fr{k\sqrt{X}}{2}-\fr{l}{12\sqrt{X}}+\fr{X_r}{12X}-\ch_r,\nonumber\\
&&H=\fr{l\sqrt{X}}{6}-\fr{X_r}{6}-X\ch_r.
\label{eq:32}
\ear
the result is in the Lanczos algebraic gauge when $\ch_r=0$.   The derivative 
of the metric $X_r$ appears in each term.   No matter what the choice of 
algebraic gauge (i.e. choice of $\ch_r$)  we cannot remove it.   Thus the 
continuity or otherwise of the Lanczos tensor depends on the continuity 
or otherwise of the derivative of the metric.   We can make any possible
discontinuity in any single component,  or even a whole component vanish by 
means of a suitable algebraic gauge.   For example choosing
\be
\ch_r=\fr{-k\sqrt{X}}{2}-\fr{l}{12\sqrt{X}}+\fr{X_r}{12X},
\label{eq:33}
\ee
gives
\ber
&&H_{trt}=0,\nonumber\\
&&H_{zrz}=k\sqrt{X},\nonumber\\
&&H_{\ph r\ph}=\fr{kX\sqrt{X}}{2}+\fr{l\sqrt{X}}{4}-\fr{X_r}{4}.
\label{eq:34}
\ear
Notice that the Lanczos tensor does not necessarily vanish for flat space-time.
For example in the Lanczos algebraic gauge or in the gauge \ref{eq:33} 
the choice
\be
k=0,~~~  l=2,
\label{eq:35}
\ee
gives flat space-time with vanishing Lanczos tensor;   however,  for example,
in the gauge $\ch_r=-X_r/6X$,  the Lanczos tensor cannot be made to vanish for 
flat space-time.   From \ref{eq:32} the metric can be expressed in terms of 
the Lanczos tensor
\be
X=(H^{~t}_{r.t}-H^{~z}_{r.z})^2/k^2.
\label{eq:36}
\ee

The differential gauge \ref{eq:9} is
\ber
&&L_{tr}=-\p_rH_{rtr}+X_r(H_{\ph t\ph}/X -H_{rtr})/2X,\nonumber\\
&&L_{r\ph}=\p_rH_{t\ph r}+X_rH_{tr\ph}/2X,\nonumber\\
&&L_{tz}=\p_rH_{tzr}+X_rH_{trz}/2X,\nonumber\\
&&L_{r\ph}=\p_rH_{r\ph r},\nonumber\\
&&L_{rz}=\p_rH_{rzr}+X_r(-H_{\ph z\ph}/X-H_{rzr}H)/2X,\nonumber\\
&&L_{\ph z}=\p_rH_{\ph zr}+X_rH_{rz\ph}/2X.
\label{eq:37}
\ear
For the Lanczos differential gauge condition \ref{eq:9}, 
\ref{eq:37} can be expressed in the integral form
\ber
&&H_{r\ph r}=a_1,\nonumber\\
&&H_{ztr}=\fr{a_2}{\sqrt{X}},\nonumber\\
&&H_{\ph tr}=\fr{a_3}{\sqrt{X}},\nonumber\\
&&H_{rtr}=\fr{1}{\sqrt{X}}(a_4+\int X_rX^{-3/2}H_{\ph t\ph}dr),\nonumber\\
&&H_{rzr}=\fr{1}{\sqrt{X}}(a_5+\int X_rX^{-3/2}H_{\ph z\ph}dr),\nonumber\\
&&H_{z\ph r}=\fr{1}{\sqrt{X}}(a_6+\int X_rX^{-3/2}H_{r\ph z}dr),
\label{eq:38}
\ear
where $a_1\ldots a_6$ are constants.   This illustrates a property of the 
differential gauge alluded to in section II;  the components of the Lanczos
tensor in \ref{eq:38} do not coincide with any of those in \ref{eq:32},  
thus these components do not participate in the 
construction of the Weyl tensor.

Here the Lanczos tensor cannot be expressed as the gradient of an 
anti-symmetric tensor of the second order.  Using that the space-time
is only $r$ dependent and that the Christoffel symbols are \ref{eq:14},  
\ref{eq:10} gives
\be
F_{tr;t}=F_{zr;z}=F_{\ph t;\ph}=F_{rt;t}=F_{rz;z}=F_{r\ph;\ph}=0.
\label{eq:39}
\ee
Any added current to \ref{eq:10} would have component $J_z$, 
and then the cyclic identity \ref{eq:5} would fail.
\section{The Covariant Derivative.}
\label{sec:IV}
In the Aharonov-Bohm effect,  the electro-magnetic field alters 
the dynamics of test particles because the electro-magnetic covariant 
derivative replaces the partial derivative in the test particles 
Schr\"odinger's equation.   
In this section we list $15$ possible covariant derivatives involving
the Lanczos tensor.   None of the possibilities can be used to complete our
analogy with the Aharonov-Bohm effect.   This is because the criteria of 
continuity cannot be used to fix the algebraic gauge in the example in
the last section,  and all the possibilities give different results depending
on algebraic gauge.   We denote the covariant derivative by $D_a$   and the
coupling constant by $c$.
\be
i)~~~\p_a\rightarrow D_a=+cH^{~b}_{a.b}.
\label{eq:40}
\ee
From ref{eq:7} and \ref{eq:8} we see immediately that this choice depends 
on the algebraic gauge.   The choice 
\be
ii)~~~\p_a\rightarrow D_a=+cH^b_{.ab},
\label{eq:41}
\ee
is the same as \ref{eq:40} with the sign of $c$ reversed,  
by virtue of the symmetry \ref{eq:4}.    The choice 
\be
iii)~~~\p_a\rightarrow D_a=+cH^b_{~ba},
\label{eq:42}
\ee
gives that the covariant and partial derivative are identical by \ref{eq:4}.
\be
iv)~~~\p_a\rightarrow  D_a=+cH^{~B}_{a.B},
\label{eq:43}
\ee
where $B$ is a fixed component not summed.   Covariant derivatives of this type
have all the disadvantages of \ref{eq:40} to \ref{eq:42},  
with the added disadvantage of picking out one component.
\be
v)~~~\p_a\rightarrow   D_a=\p_a+c(3H^{~t}_{a.t}-H^{~b}_{a.b},
\label{eq:44}
\ee
for our example this is invariant under the choice of algebraic gauge,
however by \ref{eq:8}
\be
H^{~b}_{a.b}=3\ch_r,
\label{eq:45}
\ee
thus this choice amounts to no more than an arbitrary choice of component 
in the Lanczos algebraic gauge.
\be
vi)~~~\p_a\rightarrow D_a=\p_a+c\ep_{abcd}H^{bcd}_{...},
\label{eq:46}
\ee
in our example,  or more generally in any space-time which can be expressed 
with a diagonal metric,  components of the Lanczos tensor with a identical
adjacent indices plays an essential role,  
but they would not effect this covariant derivative.
\be
vii)~~~\p_a\rightarrow D_a=\p_a+cH_{abc}p^{bc}_{..},
\label{eq:47}
\ee
where $p^{bc}_{..}$ is the stress tensor of the "test" particle.   
This coupling is unusual as the test particles own stress contributes,  
it is no longer just a test particle.   
By changing the algebraic gauge the contribution of $p^{bc}_{..}$   
changes in an arbitrary manner and thus this choice is un-useable.

   viii)Require that the covariant derivative coincides with the weak field 
covariant derivative when the fields are weak.   The weak field covariant 
derivative is
\be
2\et_{ad}W^{.d}_{b.c}=\p_bh_{ac}+\p_ch_{ab}-\p_ah_{bc}.
\label{eq:48}
\ee
Using the algebraic gauge 
\be
\ch_a=-\fr{1}{6}h_a,
\label{eq:49}
\ee
\ref{eq:12} becomes
\be
4H_{abc}=\p_bh_{ca}-\p_ah_{bc}, 
\label{eq:50}
\ee
using \ref{eq:4} this is 
\be
\p_ah_{bc}=-2H_{abc}. 
\label{eq:51}
\ee
Substituting \ref{eq:50} into \ref{eq:47} and using \ref{eq:4} and \ref{eq:5} 
gives
\be
2\et_{ad}W^{.d}_{b.c}=4H_{abc}+2H_{acb},
\label{eq:52}
\ee
now 
\be
\et_{ad}W^{~d}_{c.b}=\et_{ad}W^{~d}_{b.c}, 
\label{eq:53}
\ee
thus from \ref{eq:51} and \ref{eq:52}
\be
H_{abc}=H_{acb},
\label{eq:54}
\ee
using \ref{eq:4} and \ref{eq:5},  \ref{eq:54} gives
\be
H_{bca}=0.
\label{eq:55}
\ee
Thus the weak field covariant derivative cannot be expressed in terms of the
weak field expression for $H_{abc}$,  and we cannot require that the covariant 
derivative coincides with the weak field covariant derivative when the fields 
are weak.

   ix)Apply $H_{abc}=F_{ab;c}$ and then use the electro-magnetic covariant 
derivative.   This is too restrictive,  there is no $F_{ab}$ 
satisfying this in our example.
\be
x)~~~\p_a\rightarrow D_a=\p_a+cH_a,
\label{eq:56}
\ee
where 
\be
H_a=H_{abc}H^{bc}_{..},
\label{eq:57}
\ee
and $H^{bc}_{..}$ is given by \ref{eq:6}.   This has the disadvantage that it 
involves products and derivatives of the Lanczos tensor;  the electro-magnetic
covariant derivative has no products and derivatives of the vector potential.
For the example of the previous section in the Lanczos algebraic gauge
\be
H_r=-\fr{k^2X_r}{2}-\fr{X_rX_{rr}}{24X^2}+\fr{5X_r^3}{144X^3}
    +\fr{l^2X_r}{72X^2}-\fr{7lX_r^2}{144X^{3/2}}+\fr{lX_{rr}}{24X^{3/2}},
\label{eq:58}
\ee
in the $\ch_r=X_r/12X$ gauge
\be
H_r=-\fr{k^2X_r}{2}-\fr{X_rX_{rr}}{16X^2}+\fr{X_r^3}{32X^3}      
    +\fr{l^2X_r}{72X^2}-\fr{lX^2_r}{24X^{3/2}}+\fr{lX_{rr}}{24X^{3/2}},
\label{eq:59}
\ee
in the $\ch_r=-X_r/6X$ gauge
\be
H_r=-\fr{k^2X_r}{2}-\fr{X_rX_{rr}}{8X^2}+\fr{X_r^3}{16X^3}
    +\fr{l^2X_r}{72X^2}-\fr{lX_r^2}{X^{3/2}}+\fr{lX_{rr}}{24X^{3/2}},
\label{eq:60}
\ee
illustrating the algebraic gauge dependence of $H_a$.   

xi)Replacing ref{eq:57} by
\be
H_a=H_{bac}H^{bc}_{..},
\label{eq:61}
\ee
is the same as \ref{eq:57} with the sign of the coupling constant reversed,

xii)replacing \ref{eq:57} by
\be
H_a=H_{bac}H^{bc}_{..},
\label{eq:62}
\ee
we have that the indices $b$ and $c$ will always be identical for any 
space-time where the metric is in diagonal form and thus $H_a=0$.

Require that rather than a covariant derivative we need a change of phase
\be
xiii)~~~\al\rightarrow\al'=\al+cH_{abc}H^{abc}_{...},
\label{eq:63}
\ee
This has the same difficulties as the example of the preceeding paragraph,
and also we have a difficulty in how to integrate over the different 
trajectories.

xiv)Another possibility is to start with the Klein-Gordon equation or the 
Dirac equation in the space-time \ref{eq:13}.  
In the non-relativistic limit the Schr\"odinger equation can be recovered 
from the Klein-Gordon and Dirac equations 
and this might give information on the correct covariant derivative.

To investigate this we begin by showing that there are no solutions with
non-vanishing gravitational and Klein-Gordon field with line element 
\ref{eq:13}.   The equations for an Einstein-Klein-Gordon field are
\be
R_{ab}=2\ph_a\ph_b+2g_{ab}m^2\ph^2,   
\label{eq:64}
\ee
and 
\be
m^2\ph=\Box\ph=\fr{1}{\sqrt{-g}}(\sqrt{-g}g^{ab}_{..}\ph_a)_b.
\label{eq:65}
\ee
Thus for the line element \ref{eq:13},  using $ph_\ph=0$,
\be
R^r_{.r}=2\ph'^2+2m^2\ph^2,~~~    
R^{\ph}_{.\ph}=2m^2\ph^2,
\label{eq:66}
\ee
Now \ref{eq:15} gives $R^r_{.r}=R^\ph_{.\ph}$ therefore $\ph'=0$,  integrating
\be
\ph=\si,
\label{eq:67}
\ee
where $\si$ is a constant.  Equation \ref{eq:65} becomes
\be
m^2\ph=\fr{1}{\sqrt{X}}(\sqrt{X}\ph')',
\label{eq:68}
\ee
which vanishes by \ref{eq:67}.   Therefore $m^2=0$ and $\ph=\si$ or $\ph=0$.
Thus there are no solutions of \ref{eq:64} with line element \ref{eq:13} 
and both $m$ and $\ph$  non-vanishing.   Similarly,  using \cite{bi:PR}
it can be shown that there are no solutions of the Einstein-Dirac equations 
with line element \ref{eq:13} and 
both gravitational and Dirac fields non-vanishing.

Instead of considering coupled systems we could consider the Klein-Gordon
or Dirac fields as test fields which do not contribute to the stress of the
space-time.   Using \ref{eq:37},  \ref{eq:65}, and \ref{eq:68} the 
Klein-Gordon equation can be expressed as
\be
m^2\ph=\na_a\p^a_.=\ph"+(ln(H^{~t}_{r.r}-H^{~z}_{r.z}))'\ph',
\label{eq:69}
\ee             
suggesting the covariant derivative 
\be
xv)~~~\p_a\rightarrow D_a=\p_a+(ln(H^{~t}_{r.t}-H^{~z}_{r.z}))'.
\label{eq:70}
\ee
This covariant derivative has the disadvantage of artificially picking out a 
component and involves neither the gauge vector $\ch_a$ or the constants $k$ 
and $l$.   The analogy with the Aharonov-Bohm effect suggests that the 
exterior region should be Minkowski space-time,  where $X=r^2$, 
and that the criteria of continuity should fix the gauge vector $\ch_a$ 
and the constants $k$ and $l$;  but 
this cannot be done with covariant derivative \ref{eq:70}.
\section{Conclusion}
\label{sec:V}
An attempt was made to test in principle whether the Lanczos tensor is 
microscopically dynamically significant in the quantum realm in a similar 
manner to the vector potential.  So far the results have not produced a 
definitive result:  however whether it is possible to quantize by this method
should become clearer upon the discovery of a suitable exact solution to the 
general relativity field equations which can produce a closer analogy to the
Aharonov-Bohm experiment.

\end{document}